\documentclass[aps,preprint]{revtex4}
\usepackage{graphicx,amsmath,amssymb}

\newcommand{\bea}{\begin{eqnarray}}
\newcommand{\eea}{\end{eqnarray}}
\newcommand{\be}{\begin{equation}}
\newcommand{\ee}{\end{equation}}

\begin{document}

\begin{titlepage}
\title{Ergodicity and slow diffusion in a supercooled liquid}

\author{Neeta Bidhoodi and Shankar P. Das}
\affiliation{School of Physical Sciences,\\
Jawaharlal Nehru University,\\
New Delhi 110067, India.}

\setcounter{equation}{0}

\begin{abstract}
A model for the slow dynamics of the supercooled liquid is
formulated in terms of the standard equations of fluctuating
nonlinear hydrodynamics (FNH) with the inclusion of an extra
diffusive mode for the collective density fluctuations. If the
compressible nature of the liquid is completely ignored, this
diffusive mode sets the longest relaxation times in the supercooled
state and smooths off a possible sharp ergodicity-nonergodicity
(ENE) transition predicted in a mode coupling theory. The scenario
changes when the complete dynamics is considered with the inclusion
of $1/\rho$ nonlinearities in the FNH equations, reflecting the
compressible nature of the liquid. The latter primarily determines
the extent of slowing down in the supercooled liquid. The presence
of slow diffusive modes in the supercooled liquid do not give rise
to very long relaxation times unless the role of couplings between
density and currents in the compressible liquid is negligible.
\end{abstract}

\vspace*{.8cm}


\maketitle
\end{titlepage}

\section{Introduction}

The Mode coupling theory(MCT) has been developed as a microscopic
theory to understand the slow dynamics of a supercooled liquid. The
basic mechanism that increases the viscosity was first identified
 \cite{leth,beng,sjo-beng} from kinetic theory of dense fluids.
Subsequently equations of fluctuating nonlinear hydrodynamics (FNH)
have been used to derive \cite{mybook,kawasaki-jsp,upen-pre,prl2}
the MCT. Generally, the defining expressions for these correlation
functions are expressed in terms of space and time dependent
transport coefficients. In the FNH formulation of the MCT, it is
assumed that the crystallization process \cite{tu-frz,sjo-psica}
does not interrupt and the transport coefficients are renormalized
due to nonlinearities in the equations of motion for the slow modes
and they are expressed in terms of hydrodynamic correlation
functions. Nonlinear equations for the dynamics of the correlation
functions are obtained from such definitions combined with the self
consistent expressions for the transport coefficients. This gives
rise to a feedback mechanism \cite{beng} for slow relaxation of
correlations \cite{spd-dufty} in the supercooled liquid. As a
consequence, the mode coupling theory (in its simplest form)
predicts an ergodicty-nonergodicity (ENE) transition at a critical
density $n_c$. The correlation function $\phi(q,t)$ of collective
density fluctuations at wave vector $q$ and time separation $t$ is
$f(q)\ne{0}$ at the transition.

In Ref. \cite{DM} analysis of the fluctuating hydrodynamic equations
for the compressible liquid showed how ergodicity is restored in the
long time dynamics. The role of the $1/\rho$ \cite{yeomaj}
nonlinearities in the generalized equation for the momentum
fluctuations in the compressible fluid played the key role in
producing the ergodicity restoring mechanism. In a subsequent work
Schimitz, Dufty, and De \cite{sdd} has also considered a
self-consistent mode coupling theory for supercooled liquids.  The
analysis presented by these authors demonstrates the absence of
sharp transition to an ideal glassy phase \cite{DM} in the model. In
both the versions of mode coupling theories, respectively described
in Ref.\cite{DM} and Ref. \cite{sdd}, the density correlation has an
asymptotic behavior given by the form $[z+i \gamma(q,z)]^{-1}$,
where the kernel  $ \gamma(q,z)$ can be expressed self-consistently
in terms of hydrodynamic correlation functions giving rise to a
diffusive decay. 
Subsequent to these works several other phenomenological models
\cite{tu-kzip} for the structural relaxation in a deeply supercooled
glassy liquid \cite{sjo-tu} were developed. Models taking into
account orientational degrees of freedom \cite{tu-sd1,tu-sd2} has
been proposed. From a qualitative level the breaking of the cage
formation in the dense liquid is manifested through the couplings of
current and density fluctuations. This process influences the
dynamics and in particular mass transport in important ways.
Orientational degrees of freedom has been included in description of
the supercooled liquid to describe the process of cage formation and
freezing at a local scale. In some of the works such
phenomenological considerations were used to construct model
\cite{biman-wolynes, oppenheim1,oppenheim2,bpeak-pla} by extending
the existing self-consistent formulation of the MCT. From a general
viewpoint the effects of activated events in some of these models
are incorporated in the dynamics by using the concepts from the
random first order transition theory. The dynamic structure factor
is modified by localized activated hopping \cite{gtze-sjoren} events
termed in some works as instantons. Thus if we denote the density
correlation function which acts like an order parameter in the MCT,
as $\phi_\mathrm{MCT}(q,t)$, by including the so called hopping
\cite{rmp} it was modified as
\be \label{bw-bs}\phi(q,t)\approx
\phi_\mathrm{MCT}\phi_\mathrm{hopp} \ee
\noindent It is argued that close to the glass transition
temperature, $T_g$, since the configurational entropy $S_c$ is
diminishing, the activated process slows down leading to an arrest
of the structural relaxation. Beyond the mode coupling transition
temperature, $T_c$, the density correlation is assumed to decay via
the hopping channel. Thus the longitudinal viscosity, which is
otherwise divergent in the idealized MCT, remains finite.

In the present work we propose a model in which instead of making a
modification of standard MCT at the level of correlation function
\cite{biman-wolynes,das-schilling}, with a diffusive mode, we modify
the equations of FNH with the same, which forms the basis of the
MCT. We include an extra diffusive processes in the collective
density fluctuations in addition to the standard MCT. If the extra
diffusive mode is ignored then our model reduces to the standard
extended MCT model. The latter refers to the full mode coupling
model in which all the important nonlinearities of the original
equations of FNH are present. These nonlinearities include those
which gives rise to an ergodicity-nonergodicity (ENE) transition at
the simplest level, as well as the source of ergodicity restoring
mechanism over longer time scales. The assumed diffusive mode is an
additional mode in the system. Our analysis demonstrates the
importance of the compressible nature of the liquid in determining
the slow dynamics.


This basic model of extended MCT is obtained primarily from the
conservation laws and the dynamics of the corresponding collective
modes in the liquid. The Fluctuating Nonlinear Hydrodynamic
model\cite{sdd} is formulated in terms of two fluctuating variables
${\bf g}$ and $\rho$ and without any $1/\rho$ nonlinearity present
in the generalized equation for the momentum conservation. This
involves simplifying the expression for the kinetic energy term
$F_K$ of the driving free energy functional for the system which
determines the equilibrium state of the liquid \cite{enz-tu}. Making
this change violates the Galilean invariance of the FNH equations.
Since our focus here is primarily on the slow dynamics produced due
to dominant density fluctuations we assume that this is not too
important in the present analysis. The FNH equations studied in the
present work are also based on a purely gaussian form of the driving
free energy functional $F[\rho, g]$ like that of Ref. \cite{sdd} and
contains the same density and current coupling in the continuity
equation as in Ref. \cite{sdd}. In this model of extended MCT, the
ENE transition is smeared off due to this density and current
coupling appearing in the continuity equation. On ignoring this
nonlinearity in the continuity equation, we get the basic MCT model
which predicts an ENE transition at a critical density.  As noted
above additionally, we include here a diffusive mode as an extra
slow process in the dense supercooled state of the liquid. With this
the continuity equation is now modified and the corresponding
current have contributions from the diffusive mode as well as the
random noise. The form of a balance equation for the density
variable is maintained. The dissipative term in the equation of
motion for the collective density field $\rho({\bf x},t)$ is linear.
If this diffusive mode and the related noise is ignored then our
model reduces to the FNH model of Ref. \cite{sdd}. The goal of the
present analysis is to study how the simultaneous presence of the
diffusive mode and $1/\rho$ nonlinearities affects the dynamics and
determine their relative importance in producing the slow dynamics.

The paper is organized as follows. In the next section we discuss
the construction of the basic equations of FNH using standard
formalisms \cite{Oldbook} but adopting a purely gaussian free energy
functional. This is followed in section III  by discussion of
linearized dynamics and the noise averaged correlation functions. In
the next section VI we construct the renormalized theory taking into
account one loop expressions for the self energies renormalizing the
transport coefficients. Section V discusses the numerical solution
of the MCT equations and is followed by discussion section.

\newpage
\section{Model studied}

The basic equations of the model for the dynamics of a fluid is
obtained using the standard
techniques\cite{Oldbook,tur1-hyd,tur2-hyd} of fluctuating nonlinear
hydrodynamics (FNH). The equation of motion for the coarse grained
density $\rho({\bf x},t)$ is a continuity equation with the momentum
density ${\bf g}({\bf x},t)$ as the current which itself is a
conserved property. The current ${\bf g}({\bf x},t)$  satisfies the
momentum conservation equation. The latter constitutes the
generalized Navier-Stokes equation. We include in the present work
an additional dissipative term and a noise in the continuity
equation. The two are related by a standard fluctuation-dissipation
relation.

\subsection{Generalized Langevin Equations}

We begin with  the coarse grained mass density ${\rho}({\bf x},t)$~
and the momentum density ~${\bf g}_i({\bf x}, t)$ constitute the set
of slow variables for the liquid. In the standard formulation of
Fluctuating nonlinear hydrodynamics these variables satisfy the
Langevin dynamics with following the generalized
form\cite{ma-mazenko} for the equation of motion,
\be \label{gen-lgvn} \frac{\partial\psi_i}{\partial t} =
V_i[\psi_i]- L^0_{ij} \frac{\delta{F}}{\delta\psi_j}+\theta_i~~. \ee
In the above equation and throughout this paper we follow the
notation that repeated indices are summed over. $V_i[\psi_i]$ is the
''streaming velocity'' term representing the reversible part of the
dynamics, and is obtained as
\bea V_i({\bf x}) &=& Q_{ij} \frac{\delta{F}}{\delta {\psi_j}}
\nonumber
\end{eqnarray}
where $Q_{ij}{\equiv}\{\psi_{i},\psi_{j}\}$ is Poisson bracket
between slow variables $\psi_i$ and $\psi_j$. The driving free
energy is a functional of the slow modes in FNH description and is
written as a sum of two parts here,
\be \label{fe-t} F[\psi]= F_K [\psi]+ F_{U}[\psi] ~~.\ee
The kinetic part\cite{langer} and the potential part of free energy
functional with the variables $\{\rho,{\bf g}\}$ are respectively
given by
\bea \label{fe-k} F_K[g] &=& \int d{\bf x} ~~
\left[\frac{g^2({\bf x})}{2\rho_{0}}\right] \\
\label{fe-p} F_{U}[\rho] &=& \frac{1}{2}\int d{\bf x}~\int d{\bf
x}^{\prime}~ \delta\rho({\bf x}) \chi_{\rho\rho}^{-1}({\bf x}-{\bf
x}^{\prime})\delta\rho({\bf x}^{\prime})~~. \eea
$\chi^{-1}_{\rho\rho}$ denotes the inverse of the static (equal
time) correlation function of density fluctuations and is related to
the static structure factor\cite{hansen} of the liquid. The above
choice of the free energy functional is of a purely gaussian in both
fields $\rho$ and ${\bf g}$. The formulation of the equations of
fluctuating nonlinear hydrodynamics (FNH) is standard \cite{DM}. We
provide here a few details specific for the present model. With the
choice (\ref{fe-t})-(\ref{fe-p}) the streaming energy term for the
FNH equations ( signifying reversible dynamics) for density $\rho$
and momentum density ${\bf g}$ are different from the standard
results \cite{DM}. Nonlinearities appear in the streaming term
$V_\rho$ for the density equation, while gallelian invariance term
in the corresponding $V_{g_i}$ for momentum density $g_i$ is
different. The streaming velocities of ~$\rho$~and~${\bf g}$~ fields
are obtained as:

\bea \label{str-r} V_{\rho}(\bf x) &=& -\nabla.\left[\frac{{\bf
g}({\bf
x})\rho(\bf x)}{\rho_0}\right] \\
\label{str-g}
 V^i_g(\bf x) &=& -\rho({\bf x})\nabla_{\bf x}^ i
\frac{\delta{F_ U}}{\delta \rho ({\bf x})} -\nabla_{\bf
x}^j\left[\frac{ g_i({\bf x})g_ j({\bf x})}{\rho_0}\right ]
-\nabla_{\bf x}^i \left(\frac{g^2({\bf x})}{2\rho_0}\right)
\nonumber \\
\eea
The standard form for the free energy\cite{DM}  is one in which the
$1/\rho_0$ in the expression for $F_ K$ is replaced by a $1/\rho$
term. The corresponding $F_K$ gives a continuity equation which is
essential for the microscopic momentum conservation  with the
current density ${\bf g}$. This choice for $F_K$ also produces the
proper nonlinear term in the momentum equation needed for
maintaining the Galilean invariance in the FNH equations. The
advantage of using the present form is that the free energy
$F[g,\rho]$ remain gaussian. As a result the dissipative terms in
the dynamical equations ( $\sim \delta{F}/\delta{\psi}$ ) are linear
in the hydrodynamic variables.

The time reversal properties of dissipative terms are given by that
of the corresponding transport coefficients,
\be  L^0_{ij}(- t) =\epsilon_i\epsilon_ j
L^0_{ij}( t)~~.\ee
Here $\psi_i(-t)=\epsilon_i \psi(t)$. Thus the dissipative terms
like $L^0_{g\rho}$ are zero. We have $L^0_{\rho\rho}(- t)=
L^0_{\rho\rho}( t)$ is nonzero. Similarly $L^0_{g_ig_j}$ is also
survives the time reversal symmetry. In the present model
dissipative terms are present in both the density and momentum
equations. As a result, both equations have respective random noise
components as well. These noises in the $\rho$ and ${\bf g}$
equations are respectively denoted as ~$ f$~ and ~$\theta$. The
random part of the density equation is expressed in terms of a force
vector ${\bf f}_R$ such that ${ f}=\nabla.{\bf f}_R$. The
corresponding noise correlation functions are given by the standard
fluctuation dissipation relations.
\begin{eqnarray}
\langle f_R^i ({\bf x}, t)~ f_R^j ({\bf x}^{\prime}, t^{\prime})\rangle
&=& 2 k_B T~\delta_{ij}~L_{\rho}
\delta({\bf x}-{\bf x}^{\prime})~\delta( t- t^{\prime})\nonumber\\
\langle \theta_ i ({\bf x}, t)~\theta_ j({\bf x}^{\prime},
t^{\prime})\rangle &=& 2  k_{B} T~ L^0_{ij}\delta({\bf x}-{\bf
x}^{\prime})~\delta( t-t^{\prime})\nonumber
\end{eqnarray}
where we have taken the dissipative coefficient in the density
equation as in the equation as $L^0_{\rho\rho} = {L_\rho}
\nabla^{2}$.

Using the Poisson brackets between the slow variables\cite{volvick}
described in the appendix \ref{appendix-A}, the generalized Langevin
equation of motion for~$\rho$~ having both dissipative and random
parts is obtained as
\be \frac{\partial \rho}{\partial t}+\nabla . \left ({\bf g}({\bf
x})\frac{\rho}{\rho_0} \right ) -\gamma_0\nabla^2{\rho ({\bf x})} =
f~~, \ee
where $\gamma_0={L_\rho}\chi^{-1}_{\rho\rho}$. The above equation is
written in the form of a continuity equation as
\be \label{cont-eqn}
 \frac{\partial \rho}{\partial  t}+\nabla
.\tilde{\bf g}=0~~,\ee
with the generalized current $\tilde{\bf g}(\bf x,t)$ obtained as
\be \label{def-tilg} \tilde{\bf g}= {\bf
g}+\frac{\delta\rho}{\rho_0}{\bf g} -\gamma_0{\bf \nabla} \rho -{\bf
f}_R ~~.\ee
which is different from ${\bf g}(\bf x,t)$ appearing in the first
term on the right hand side. The field ${\bf g}$ is such that its
dynamics is described by the generalized Langevin equation
(\ref{gen-lgvn}), and the reversible part of its dynamics being
given in terms Poisson brackets for the microscopic field ${\bf g}$.
The generalized Langevin equation of motion for~${\bf g}$~ field is
given as
\bea \label{mom-eom} &&\frac{\partial g_ i}{\partial
t}+\nabla_j\left[\frac{g_ i g_j}{\rho_0}\right] +\nabla_i\left
[\frac{g_i^2} {2\rho_0}\right ] + \rho\nabla_i \frac{\delta
F_U}{\delta \rho} +{\tilde L}^0_{ij}{g_j}=\theta_i~~. \eea
In a normal fluid the diffusive mode is absent and the two
quantities ${\bf g}$ and $\tilde{\bf g}$ are the same. We have
defined the matrix of kinematic viscosity coefficients in terms of
$\tilde{ L}^0_{ij}= L^0_{ij}/{\rho_{0}}$ and the nonzero bare
transport matrix elements are obtained as
\be L^0_{ij} = -\eta_0[\frac{1}{3}\nabla_
i\nabla_j+\delta_{ij}\nabla^2] -\zeta_0\nabla_
i\nabla_j~~. \ee
$\zeta_0$ and $\eta_0$ respectively denotes the bulk and shear
viscosities of the liquid. Equation (\ref{mom-eom}) does not imply
microscopic conservation for the total momentum current $\tilde{\bf
g}$ appearing in the continuity equation.

As a result of using the purely gaussian free energy functional
there are several changes in the equations of motion:

\noindent (a) The convective nonlinearity of the standard form
$\nabla_j[g_ig_j/\rho]$ which is essential for Galilean invariance
is now absent from the equation for the momentum density {\bf g};

\noindent (b) The equations of motion must have the Poission bracket
between slow variables unchanged and the detailed balance condition
$Q_{ij}=-Q_{ji}$ is not compromised.

\noindent As a result the continuity equation now contain extra
density-momentum nonlinearities. These additional bi-linearities
eliminate the structural arrest predicted by mode coupling theories
with only density-density nonlinearities in the Pressure term
\cite{das-PRL84}. Interestingly, the form of the cutoff function in
this model, responsible for the removal of the sharp transition, is
identical at the one loop order to the same quantity in the
 analysis presented in Ref. \cite{das-PRE96}. This holds even when the
diffusive mode in collective density fluctuations is ignored. In the
next section we consider the implications of the full nonlinear
model with both the diffusive mode and density current
nonlinearities being present.

The issue of the microscopic momentum conservation in the present
model is special. The FNH equations considered here are plausible
generalizations of the long time, long length scale hydrodynamics.
The continuity equation is maintained in the present model at the
microscopic level, {\em i.e.},  with the microscopic definition of
density $\hat{\rho}$ in terms of delta functions we obtain a
continuity equation with a corresponding microscopic momentum
density $\hat{\bf g}$. On averaging the microscopic continuity
equation we obtain a similar equation (\ref{cont-eqn}) involving the
coarse grained densities. The coarse grained current is $<\hat{\bf
g}({\bf x},t)>{\equiv}\tilde{\bf g}({\bf x},t)$, defined in the Eqn.
(\ref{def-tilg}). Note that the current ${\bf g}({\bf x},t)$ which
satisfies the balance Eqn. (\ref{mom-eom}) is different from the
coarse grained current $\tilde{\bf g}$. With this interpretation,
the flux in the continuity equation for density $\rho({\bf x},t)$,
{\em i.e.}, $\tilde{\bf g}$ does not follow a balance equation and
hence the momentum conservation is not preserved. On the other hand,
if we take ${\bf g}$ as the coarse grained momentum density current,
the continuity equation and Gallelian invariance are both violated.
In systems with microscopic Brownian dynamics for which the mode
coupling models, such as the present one, are often applied,
momentum conservation is not satisfied though not implied in the
strict microscopic sense. For the frozen amorphous state, the
momentum fluctuations decay out much faster compared to the density
fluctuations. Thus for the decay of density fluctuations
approximations without strict momentum conservation is assumed.
Works of Kawasaki\cite{kawa-miya,munakata} obtaining the
Dean-Kawasaki equations\cite{dean} using the so called adiabatic or
over-damping approximation of the momentum equation are similar in
this respect. Even in systems with Newtonian dynamics this
approximation is applied for studying glassy behavior using density
as the only relevant collective variable.

\subsection{Correlation and Response functions}

We use the standard field theoretic method of Martin-Siggia-Rose
(MSR)\cite{msr,jensen,wagner,hca1,hca2,hca3} in order to obtain the
perturbative corrections due to the various nonlinearities. The full
matrix ${\bf G}$ of correlation between the various fields,
respectively at two different space time points (denoted as 1 and 2)
includes the correlation functions and the response functions. These
are respectively defined as
\bea \label{cor-fn} G_{\alpha\beta}(12) &=& \langle\psi_{\beta}
(2)\psi_{\alpha}(1)\rangle \\
\label{res-fn} G_{\alpha\hat{\beta}}(12) &=&
\langle\hat{\psi}_{\beta} (2)\psi_{\alpha}(1)\rangle
\eea
The Greek letter subscripts refer to the set of physical fields
$\{\psi\}\equiv\{\rho,{\bf g}\}$  and their respective hatted
counterparts $\hat{\psi}\equiv\{\hat{\rho},\hat{\bf g}\}$. The
averages are functional integrals over all the fields weighted by
$\exp [-{\cal A}]$ where the action functional ${\cal A}$ is
obtained\cite{DM} as \be {\cal A}[\psi,\hat{\psi}] \equiv{\cal
A}_0[\psi,\hat{\psi}]+{\cal A}_I[\psi,\hat{\psi}]\ee using the
generalized Langevin equations. ${\cal A}_0$ and ${\cal A}_I$ denote
the gaussian (quadratic in the fields) and nongaussian parts
originating respectively from the linear and nonlinear parts of the
equations of motion for $\{\rho,{\bf g}\}$. Using the equations of
motion for the $\rho$ and ${\bf g}$ fields we obtain the MSR action
functional involving the conjugate fields $\{\hat{\rho},\hat{\bf
g}\}$ of the MSR approach\cite{mybook} in the following form.
\bea \label{MSR-action} {\cal A}[\psi,\hat \psi] &=& \int d1 \Bigg [
\beta^{-1}\Big
\{\hat\rho(1)\gamma_0\chi_{\rho\rho}\nabla^2\hat\rho(1) + \hat {
g}_i(1) L^{0}_{ij}(1)\hat g_
j(1) \Big \}\\
&+& i\hat{\rho}(1)\Big \{ \Big ( \frac{\partial \rho(1)}{\partial
t}+\nabla.{\bf g}(1)-\gamma_0\nabla^{2}\rho(1)\Big )+
\nabla.\Big ( {\bf g}(1)\frac{\delta\rho(1)}{\rho_0}\Big ) \Big \} \nonumber \\
+ i \hat {g}_i(1) \Big \{ \frac{\partial
g_i(1)}{\partial t} &+& \rho(1)\nabla_i
\chi^{-1}_{\rho\rho}{\delta\rho(1)} + {\tilde L}^0_{ij}{
g}_j(1) + \nabla_j \left[\frac{g_i(1)g_
j(1)}{\rho_0}+ \delta_{ij}\frac{g_
j^2(1)}{2\rho_0}\right] \Big \} \Bigg ] \nonumber \\
&\equiv& {\cal A}_0[\psi,\hat{\psi}]+{\cal A}_I[\psi,\hat{\psi}]
\eea
In writing the above form of the action we have used the equation of
motion (\ref{mom-eom}) corresponding to the Gaussian form of the
driving free energy functional (\ref{fe-p}). The inverse of zeroth
order matrix ~$G_{\alpha\beta}^{0}$~ corresponding to the gaussian
part ${\cal A}_0$ of the above action functional (\ref{MSR-action})
is given in table \ref{table-G0}.

\subsection{Fluctuation-Dissipation Relations}

The transformations which keep the MSR action invariant are written
in terms of the field $\psi_i$ as
\bea \psi_i({\bf x}, - t)&\rightarrow& \epsilon_
i\psi_i({\bf x}, t)\\
\hat{\psi}_ i({\bf x}, - t)&\rightarrow& -\epsilon_ i
\left[\hat{\psi}_ i({\bf x}, t)- i \beta \frac{\delta
F}{\delta\psi_ i({\bf x}, t)}\right],~~\eea
For example, we use the above transformation to obtain the FDT's.
Using the transformation for the $\rho$ field
\bea \rho({\bf x}, - t) &\rightarrow& \rho({\bf x}, t) \\
\hat{\rho}({\bf x}, - t) &\rightarrow& - \left[\hat{\rho}({\bf x},
t)- i \beta \frac{\delta F}{\delta\rho({\bf x},
t)}\right]~~. \eea
Using these time reversal invariance properties\cite{ABL,DM} of the
action ${\cal A}$, a set of fluctuation-dissipation relations (FDR)
linking the correlation and response functions is obtained.
\bea \label{FDR-g}  G_{g_ j \beta}(q,\omega) &=&
-2\beta^{-1}\rho_0 {\rm  Im} [ G_{ \hat g_ j
\beta}( q,\omega)]\\
\label{FDR-r} G_{ \rho \beta}(q,\omega) &=&
-2\beta^{-1}\chi_{\rho\rho} {\rm Im }[G_{ \hat{\rho}
\beta}(q,\omega)]~~, \eea
where $\beta$ is an un-hatted variable. This model have a complete
set of FDR linearly relating correlation and response functions.
This is a consequence of having gaussian free energy functional
\cite{ABL,DM09,jsp2-ene}. To summarize, using the time translational
invariance properties of the action (\ref{MSR-action}), the
fluctuation dissipation relation between correlation and response
functions involving the field $\hat{\psi}$ and $\varphi$ are
obtained in the form :
\be \label{fdr-reln} G_{\zeta_\psi{\varphi}}({\bf q},\omega) =
-2\beta^{-1}{\rm Im} G_{\hat{\psi}\varphi}({\bf q},\omega),\ee
where $\zeta_\psi$ is expressed as a functional derivative of the
free energy functional with the field $\psi$:
\be \label{zeta-def} \zeta_\psi({\bf x})= \frac{\delta F}{\delta
\psi({\bf x})} ~~.\ee
In the present case since $F$ is quadratic in the fields
$\{\rho,{\bf g}\}$, the function $\zeta$ is linear in the fields.
Hence the resulting FDT's are therefore linear\cite{mazenko-book}.

\subsection{Renormalized Dynamics}

In this section we discuss how the nonlinearities in the FNH
equations renormalize the dynamics using standard field theoretic
techniques. The role of the nonlinearities in the dynamics is
expressed in terms of the self energy matrix $\Sigma$ defined
through the Dyson equation
\be \label{dyson}
 {\bf G}^{-1}(1,2)={\bf G}^{-1}_{0}(1,2)-\Sigma(1,2),~~\ee
where $G_0^{-1}$ and $G^{-1}$ respectively denote the inverse of the
correlation matrices obtained with gaussian action ${\cal A}_0$ and
full action ${\cal A}$. The key quantity which determined the
renormalization of the correlation function matrix from ${\bf G}_0$
to ${\bf G}$, is the self energy matrix $\Sigma$.

In the Appendix \ref{appendix-A} we give a brief description of the
structure of the Green's function matrix in this problem. We
demonstrate that by inverting the matrix ${\bf G}^{-1}$, the
correlation functions of collective modes are obtained in a form
which is real by construction and involves the response type
correlations between hatted and un-hatted fields. The renormalized
correlation and response functions are now expressed in terms of
renormalized transport coefficients as shown in Eqn.
(\ref{ren-longa}). The renormalization of the transport coefficients
in the model due to the nonlinearities in the equations of motion
for the collective modes is a key ingredient in the present
analysis. We briefly outline the details of obtaining the
expressions for correlation functions (\ref{eq:84a}) and response
 functions (\ref{resp-funa}) using the MSR formalism in the Appendix
\ref{appendix-A}. The renormalizability is demonstrated here in the
hydrodynamic limit. To understand this we need to analyze the nature
of the renormalized theory in the present case. We note the
following points specific to the present model and essential for
further analysis.

\begin{itemize}

\item[(i)]
The FDR relations between the correlation ($G_{\psi\psi}$) and
response functions ($G_{\psi\hat{\psi}}$) of the MSR theory
presented here are linear.

\item[(ii)]The Dyson equation (\ref{dyson}) links {\em inverse} of ${\bf G}$
and ${\bf G}_0$. Effects of nonlinearities are accounted for through
the self energy matrix $\Sigma$. The elements of the zeroth order
Green's functions ${\bf G}^{-1}_0$ involve only the bare transport
coefficients. The corresponding full ${\bf G}$ for the nonlinear
dynamics involve the respective renormalized transport coefficients.

\item[(iii)]The matrices ${\bf G}_0^{-1}$ and  ${\bf G}^{-1}$ both have
correlation ($\psi\psi$) and response ($\psi\hat{\psi}$) elements
which are expressed in terms of transport coefficients. For ${\bf
G}^{-1}_0$ it is only {\em bare} transport quantities. For ${\bf
G}^{-1}$, the corresponding self energy elements appearing
respectively in its correlation and response elements must be linked
to each other ensure that the renormalized transport coefficients
are unique. However the full $G^{-1}$ matrix is not simply obtained
from the $G_0^{-1}$ with bare quantities being replaced by
renormalized transport coefficients. Some of the elements of
$\Sigma$ are nonzero though the corresponding element for ${\bf
G}_0^{-1}$ matrix are zero. Thus for the full theory, a linear FDR
between the elements of ${\bf G}$ does not translate in to a similar
relation between the corresponding elements of $\Sigma$. Similar
relations we have established here hold only in the hydrodynamic
limit.

\end{itemize}

\section{Dynamics of Correlation functions}

\begin{figure}[!htb]
\includegraphics*[width=0.3\textwidth]{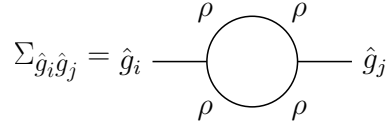}
\centering \caption{One loop contributions to $\Sigma_{{\hat
g}_i{\hat g}_j}$.}\label{fig01}
\end{figure}

\begin{figure}[!htb]
\includegraphics*[width=0.5\textwidth]{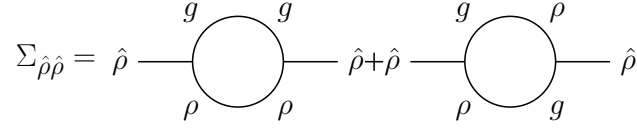}
\centering \caption{One loop contributions to
$\Sigma_{\hat\rho\hat\rho}$.}\label{fig02}
\end{figure}

Using the linear FDR relactions (\ref{FDR-g})-(\ref{FDR-r}) we
obtain that the Laplace transformation of three correlation
functions are
\bea \label{crr} {\cal
D}\psi_{\rho\rho}(q,z)&=&{z+iq^2L(q,z)} \\
\label{cgg} {\cal D}\psi_{ g g}(q,z)&=&{z+iq^2\gamma(q,z)}\\
\label{crg} {\cal D}\psi_{\rho  g}(q,z)&=& -iqc
\eea
where the denominator ${\cal D}$ has been defined in Eqn.
(\ref{detrmina}) in the Appendix \ref{appendix-A}. From the above
relations we obtain the density correlation function as,
\be \label{grr} \frac{1-z\psi_{\rho\rho}}{\psi_{\rho\rho}}+
\frac{q^2c^2}{z+iq^2L(q,z)}=iq^2\gamma(q,z)~~. \ee
The term $\gamma$ appearing on the right hand side of the above
equation is the sum of the contributions from the self energy
$\gamma_{\hat{\rho}\rho}$ and the bare diffusion process $\gamma_0$
introduced here. The self energy contribution arises from the
bilinear couplings of density and momentum in the continuity
equation (\ref{cont-eqn}). In absence of this coupling, the self
energy with $\gamma_{\hat{\rho}\rho}$ is zero. If both the bare
diffusion and density-current couplings are absent, the continuity
equation has the flux ${\bf g}$ and we obtain the standard form of
the density correlation function.
\bea
\psi_{\rho\rho}(q,z)&=&\frac{z+iq^2L(q,z)}{z[z+iq^2L(q,z)]-q^2c^2}\nonumber\\
\psi_{\rho {g}}(q,z)&=&\frac{-iqc}{z+iq^2L(q,z)}\psi_{\rho\rho}(q,z)
\eea
By doing an inverse Laplace transform of the Eqns.
(\ref{crr})-(\ref{crg}), a set of integro-differential equations for
the time evolution of the correlation functions
$\{\psi_{\rho\rho}(q,t),\psi_{\rho{g}}(q,t),\psi_{g g}(q,t)\}$ are
obtained. A fully wave vector dependent solution of the problem is
very involved. We focus here on the basic features of the mode
coupling dynamics, suppressing all wave vector dependence. In the
schematic form of the model, wave vector dependence of the
correlation functions are dropped obtaining the following set of
coupled equations for the time dependent correlation functions:
\bea \label{grr-t} &&\ddot{\psi}_{\rho\rho}(t) + \kappa^2 \Bigg [
\psi_{\rho\rho}(t)+\gamma(t) + \int_{0}^{t} d\tau \left [
L(t-\tau)+\gamma(t-\tau) \right ]\dot{\psi}_{\rho\rho}(\tau)\nonumber\\
&+& \kappa^2\int_{0}^{t} d\tau \gamma(t-\tau)\int_0^{\tau}
{d\tau'}\Big [ L(\tau-\tau')\psi_{\rho\rho}(\tau') \Big ] \Bigg ]
=0~~\\
\label{ggg-t} &&\dot{\psi}_{ g g}(t)=\dot{\psi}_{\rho\rho}(t)-
\kappa^2\int_{0}^{t} d\tau \left [ L(t-\tau){\psi}_{g
 g}(\tau)-\gamma(t-\tau){\psi}_{\rho\rho}(\tau) \right ] \nonumber \\
\label{grg-t} &&\dot{\psi}_{\rho {g}}(t)=-\kappa
~\psi_{\rho\rho}(t)-\kappa^2\int_{0}^{t} d\tau L(t-\tau){\psi}_{\rho
{g}}(\tau) ~~.\eea
In the above equation we have denoted the wave vector $q$ scaled
with an upper cutoff $\Lambda$ as $\kappa=q/\Lambda$. Time has been
rescaled in units of $\tilde{L}_0/c^2$ involving the bare
(kinematic) longitudinal viscosity $\tilde{L}_0=L_0/\rho_0$. Hence
frequency is scaled in units of $c^2/\tilde{L}_0$. We also take the
dimensionless quantity $(\Lambda\tilde{L}_0/c)=1$.  Assuming wave
vector independent structure function $\chi$ (say), the mode
coupling integrals are expressed in terms of the dimensionless
coupling constant $\lambda=\Lambda^3/(6\pi^2{c^2}\beta\rho_0)$. In
the schematic model all wave vector dependence of the theory is
incorporated through the parameter $\lambda$. To properly account
for the structural effect, this dependence should be replaced by
that of the static (equal time) correlation functions. The latter
depend on the thermodynamic parameters for the liquid state. The
renormalized memory functions are obtained as a sum of the bare and
mode coupling parts,
\bea
L(t)&=&\delta(t)+L^{mc}(t)\\
\gamma(t)&=&\Delta_0\delta(t)+\gamma^{mc}(t)~~, \eea
where the constant $\Delta_0$ is proportional to the diffusion
constant $\gamma_0$ corresponding to the slow mode introduced in the
continuity equation. In terms of the bare longitudinal viscosity
$\tilde{L}_0$ of the liquid we define
\be \label{inst} \Delta_0=\frac{\gamma_0}{\tilde{L}_0} \ee
$L^{mc}(t)$ and $\gamma^{mc}(t)$ are both nonlinear functional of
$\psi_{\rho\rho}(t)$, ${\psi}_{\rho{g}}(t)$, and ${\psi}_{g g}(t)$.
At one loop level we obtain these quantities from a diagrammatic
calculation of the memory functions, dropping all wave vector
dependence. The corresponding one loop diagrams for the self
energies $\gamma_{\hat{g}\hat{g}}$ and
$\gamma_{\hat{\rho}\hat{\rho}}$ respectively representing the
renormalization of the viscosity and the ergodicity restoring
mechanisms, are shown in Fig. \ref{fig01}-\ref{fig02}. The memory
functions $L^{mc}$ and $\gamma^{mc}$ are respectively obtained as
\bea \label{mem-L}
L^{mc}(t)&=&\lambda  \psi_{\rho\rho}^2(t)\\
\label{mem-g}
\gamma^{mc}(t)&=&\frac{\lambda}{3}~\Big[\psi_{\rho\rho}(t)\psi_{ g
g}(t) +\psi_{\rho {g}}(t)\psi_{ g \rho}(t)\Big]~~. \eea
The definitions (\ref{mem-L})-(\ref{mem-g}) form a closed set of
nonlinear equations (\ref{grr-t})-(\ref{ggg-t}) for the correlation
function. If we set the extra diffusive mode of $\gamma_0$ to be
zero then this model is same as what was proposed in Ref.
\cite{sdd}. We refer this as SDD model and the corresponding
$\gamma^{mc}$ as $\gamma_{SDD}$.
\be \label{g_SDD} \gamma(t) = \gamma^{mc}(t)= \gamma_{SDD}~~. \ee
In the following we also consider the case in which we obtain
$\gamma^{mc}$ to leading order in $\lambda$ or the density-momentum
nonlinearity of the continuity equation. In this approach, we
replace the correlation function $\psi_{\rho{g}}$ and $\psi_{g g}$
in terms of derivatives of density correlation function
$\psi_{\rho\rho}$ at lowest order in the perturbation theory ( in
terms of the ${\bf g}\delta\rho$ coupling). This will be the one
loop approximation for the function $\gamma^{mc}(t)$. In presence of
the intrinsic diffusive mode $\Delta_0$ the $\gamma$ denoted as
$\gamma_{DM}$.
\be \label{mem-gDM} \gamma^{mc}(t){\equiv}=\gamma_{DM}={2\lambda}
\Big[\dot{\psi}_{\rho\rho}^2(t)
-\frac{2\Delta_0}{3}\dot{\psi}_{\rho\rho}(t)\psi(t)
-\frac{\Delta^2_0}{5}\psi_{\rho\rho}^2(t)\Big]{\equiv}\gamma_{DM}.
\ee
$\dot{\psi}_{\rho\rho}(t)$ above refers to derivative with respect
to time and so on. The decay of the correlation functions with both
$\gamma_{DM}$ and $\gamma_{SDD}$ will be presented in the following
section.

\section{Numerical Results}
In order to analyze the implications on the ENE transition we focus
on the above integro differential equation
(\ref{grr-t})-(\ref{ggg-t}). Numerical solution of the above set of
schematic equations for the correlation function provides insight in
to the nature of the dynamics under various circumstances.  In the
extended mode coupling model considered here there are clearly two
mechanisms competing with each other in eliminating the sharp ENE
transition. These are

\noindent A. The bilinear couplings of density and current ( ${\bf
g}\delta\rho$) in the compressible liquid  present in the continuity
equation similar to Ref. \cite{sdd}. This is represented by the
${\bf \nabla} \cdot [({\bf g}/\rho) \delta\rho]$ term in the
continuity equation (\ref{cont-eqn}).

\noindent B. The presence of the slow mode with diffusion constant
$\gamma_0$ for density fluctuations.

\begin{figure}[!htb]
\includegraphics*[width=0.5\textwidth]{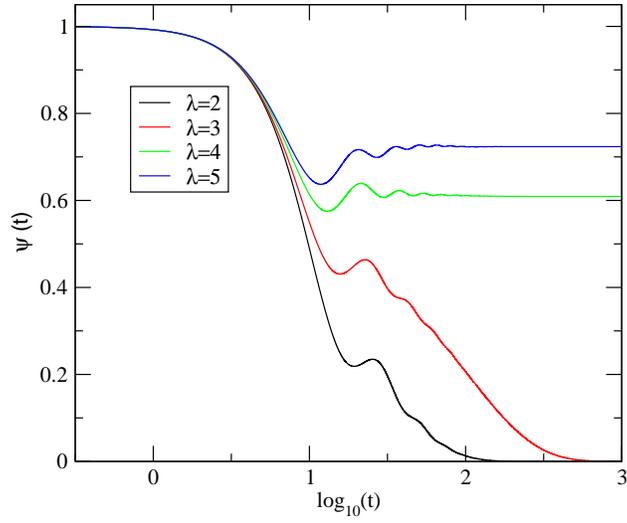}
\centering \caption{The density correlation function $\psi(t)$ Vs.
log(t) for different values of the thermodynamic parameter $\lambda$
(see text) with a sharp ENE transition at $\lambda=4$. Here all
ergodicity restoring process have been ignored} \label{fig03}
\end{figure}

\noindent If we remove both processes A and B,  by ignoring the
density momentum coupling int the continuity equation and setting
$\gamma_0=0$, then the model is the simple MCT model \cite{leth}
with an ideal transition. In Fig. \ref{fig03} the results for the
density correlation function $\psi_{\rho\rho}(t)$ for this simple
case is shown for different values of the parameter $\lambda$. The
liquid undergoes an ideal ENE transition at the point
$\lambda=\lambda_c=4$ in the model. With increasing $\lambda$ the
density correlation function relaxes slower and eventually freezes
at a nonzero value beyond $\lambda=4$.

\subsection{Dynamics of SDD Model}

\begin{figure}[!htb]
\includegraphics*[width=0.5\textwidth]{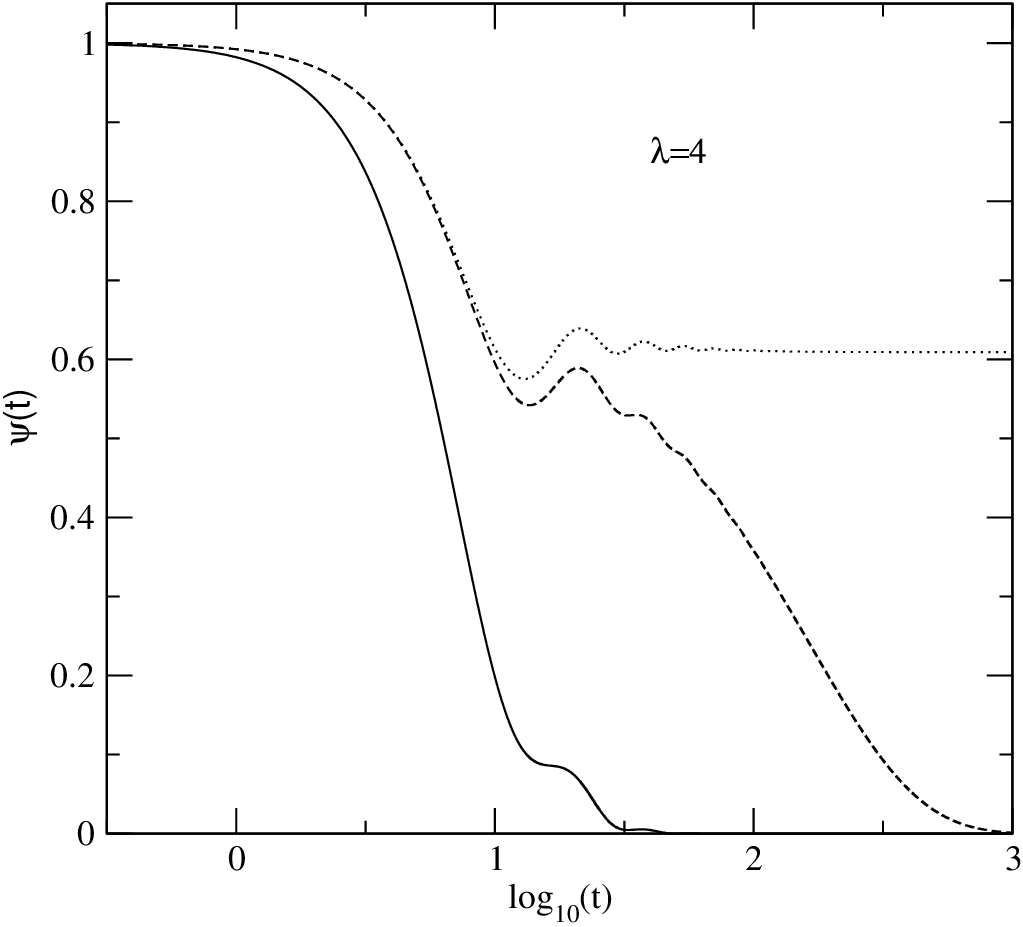}
\centering \caption{The density correlation function $\psi(t)$ Vs.
log(t) for the thermodynamic parameter $\lambda=4$ and $\gamma_0=0$.
With the cutoff function $\gamma^{mc}$(see text) set to a)
zero(dotted), b) $\gamma_{SDD}$(solid), and c) $\gamma_{DM}$
(dashed).} \label{fig04}
\end{figure}

\begin{figure}[!htb]
\includegraphics*[width=0.5\textwidth]{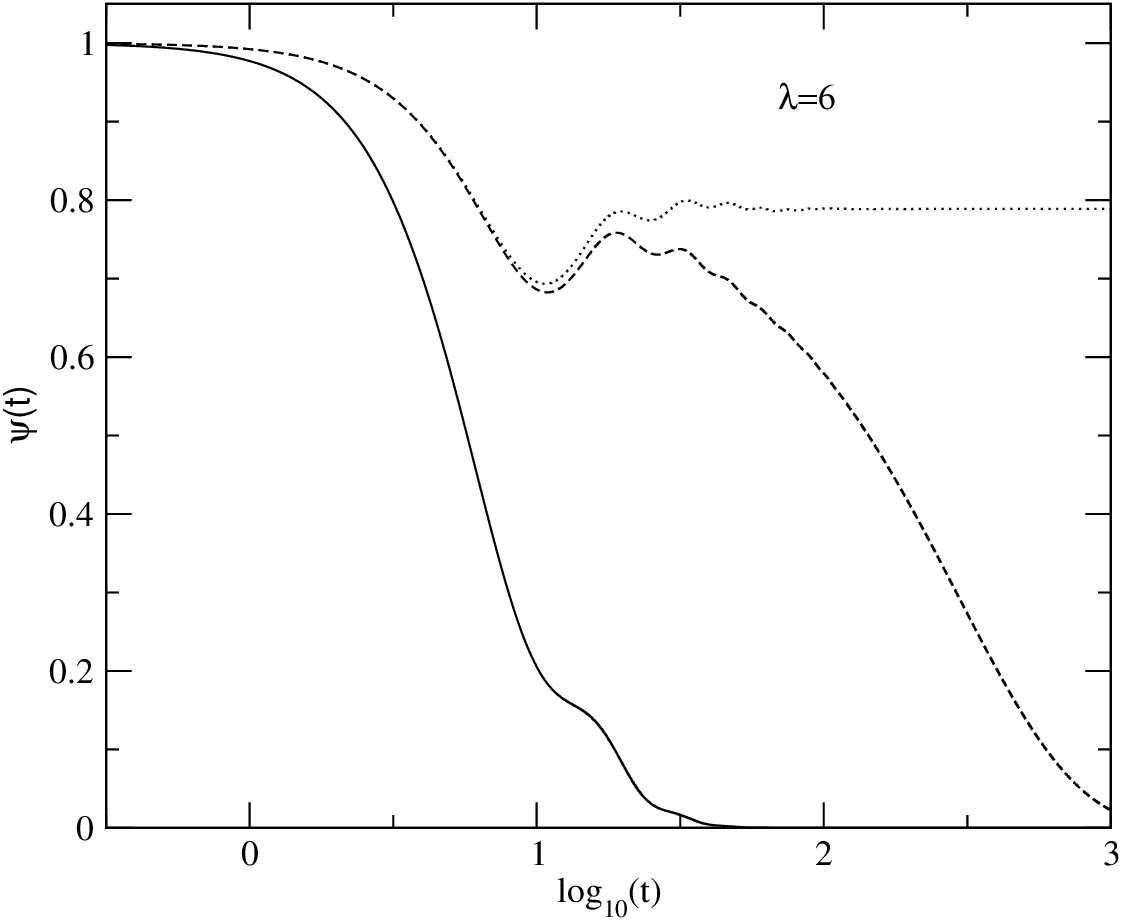}
\centering \caption{The density correlation function $\psi(t)$ Vs.
log(t) for the thermodynamic parameter $\lambda=6$ and $\Delta_0=0$.
With the cutoff function $\gamma^{mc}$(see text) set to a)
zero(dotted), b) $\gamma_{SDD}$(solid), and c) $\gamma_{DM}$
(dashed).} \label{fig05}
\end{figure}

\begin{figure}[!htb]
\includegraphics*[width=0.5\textwidth]{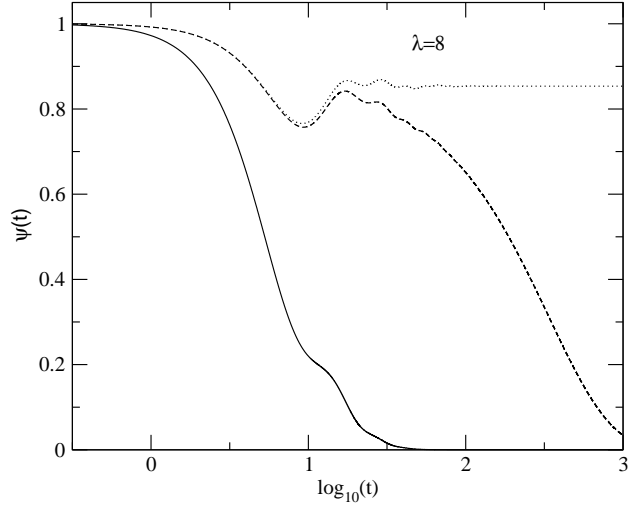}
\centering \caption{The density correlation function $\psi(t)$ Vs.
log(t) for the thermodynamic parameter $\lambda=8$ and $\Delta_0=0$.
With the cutoff function $\gamma^{mc}$(see text) set to a)
zero(dotted), b) $\gamma_{SDD}$(solid), and c) $\gamma_{DM}$
(dashed).} \label{fig06}
\end{figure}

To consider the extended model without the ENE transition we first
take the case in which process A is included while the diffusive
process of B is absent. With this choice ($\Delta_0=0$) the present
model becomes identical to what was earlier considered in Ref.
\cite{sdd}. It is well known\cite{sdd,das-PRE96} that the inclusion
of bilinear couplings in the density and momentum in the continuity
equation produce the cut off mechanism responsible for the
restoration of the ergodicity in this model. The corresponding cut
off function $\gamma^{mc}(t)$ is now obtained either in the form
$\gamma_{SDD}$ or $\gamma_{DM}$ stated above. The mode coupling
effects are considered in the expressions (\ref{g_SDD}) and
(\ref{mem-gDM}) at the one loop level. In both cases due to
contributions coming from the density and current couplings,
ergodicity is restored in the final decay of the correlation
function. In terms of the density dependent parameter $\lambda$ we
obtain the decay of the correlation functions. In Fig
\ref{fig04}-\ref{fig06}, we show the density correlations
respectively for the parameter values $\lambda=4,$ $6$, and $8$. In
each of these figures,  we display the correlation functions
obtained corresponding to different choices for the cut off function
$\gamma^{mc}$ : set equal to a) zero (dotted), b) $\gamma_{SDD}$
(solid), and c) $\gamma_{DM}$ (dashed).

\begin{figure}[!htb]
\includegraphics*[width=0.5\textwidth]{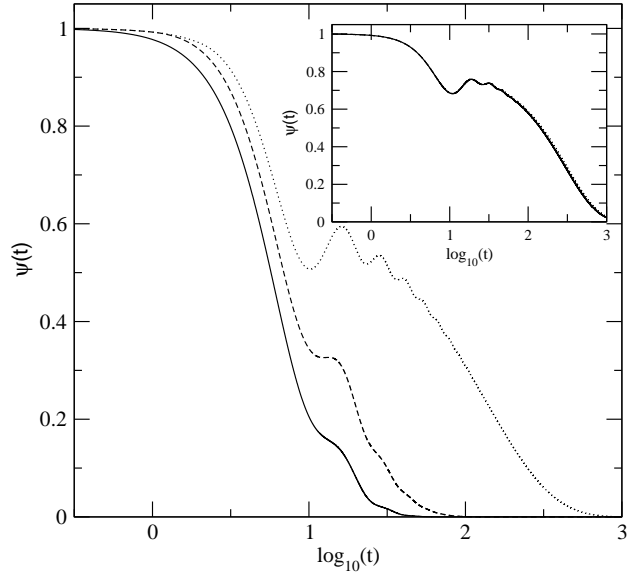}
\centering \caption{The density correlation function $\psi(t)$ Vs.
log(t) for the thermodynamic parameter $\lambda=4$ and $\Delta_0=0$.
With the cutoff function $\gamma_{SDD}$ set to start from a time
$t_0=0$(solid), $1$(dashed), and $2$(dotted). Inset shows the same
results with the corresponding cutoff function as $\gamma_{DM}$}
\label{fig07}
\end{figure}

\begin{figure}[!htb]
\includegraphics*[width=0.5\textwidth]{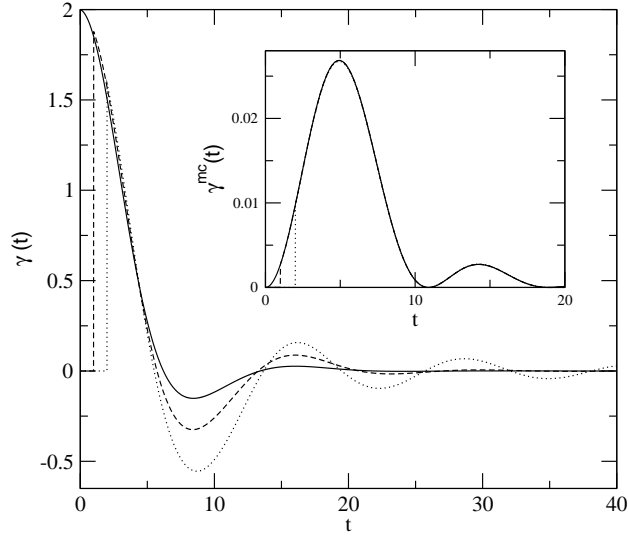}
\centering \caption{The cutoff function $\gamma^{mc}(t)$ Vs. $t$ for
the results shown in Fig \ref{fig07} in the main and inset
respectively.}\label{fig08}
\end{figure}

Another approach to understanding slow dynamics involves assuming
that the ergodicity restoring mechanisms are effective beyond an
initial time. The idea is to introduce a lower cutoff time $t_0$
from when the cutoff process is assumed to become effective
\cite{henk-t0,jcp-t0}. The role of $t_0$ on the dynamics of the
density correlation function is explored in Fig. \ref{fig07}. The
main figure shows the results for $\psi_{\rho\rho}(t)$ vs. $t$ with
$\gamma^{mc}{\equiv}\gamma_{SDD}$ for three different choices of
$t_0=0,1,$ and $2$. With $\gamma^{mc}$ being determined from
$\gamma_{DM}$ the effect of $t_0$ is less on the long time dynamics
of $\psi_{\rho\rho}(t)$ as shown in the Inset of Fig. \ref{fig07}.
The corresponding cutoff functions $\gamma_{SDD}(t)$ and
$\gamma_{DM}(t)$ are displayed in Fig. \ref{fig08}.

\subsection{Inclusion of a diffusive mode}

\begin{figure}[!htb]
\includegraphics*[width=0.5\textwidth]{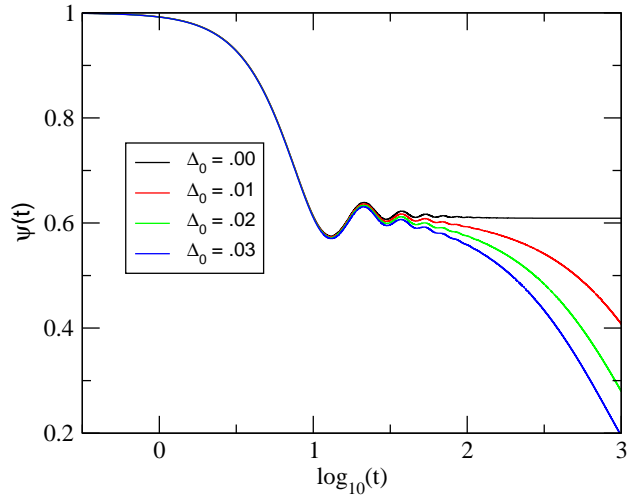}
\centering \caption{The density correlation function $\psi(t)$ Vs.
log(t) for different diffusion coefficients $\Delta_0$. The
density-current coupling of Ref. \cite{sdd} is absent here and
$\lambda=4$.} \label{fig09}
\end{figure}

Next, we consider the case in which the process A is excluded but
the diffusive mode of B is present. This means that the bilinear
density and momentum coupling in the continuity equation is ignored
but the extra diffusive mode is included in the theory taking
$\gamma_0$ nonzero. The result of this model is shown in Fig.
\ref{fig09}. We plot density correlation function $\psi_{\rho\rho}$
vs. $t$ for different choices of the relative bare diffusion
coefficients $\Delta_0$ (defined in Eqn. (\ref{inst}). With increase
of $\Delta_0$, the density correlation function decays faster. The
value of $\lambda$ used in this figure is kept constant at
$\lambda_c=4$.

\begin{figure}[!htb]
\includegraphics*[width=0.5\textwidth]{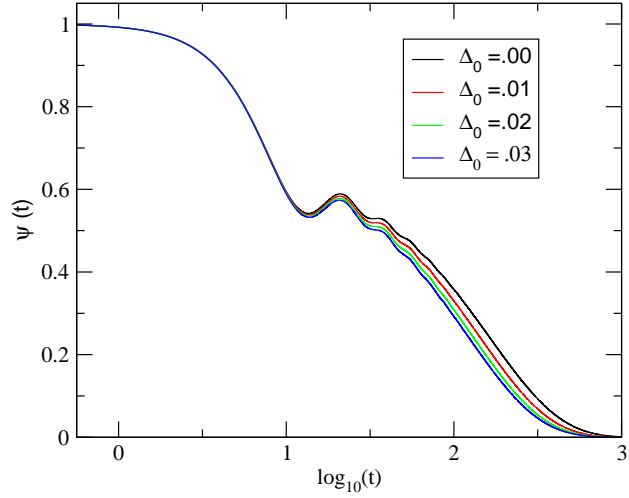}
\centering \caption{The density correlation function $\psi(t)$ Vs.
log(t) for different diffusion coefficients $\Delta_0$ together with
the density-current coupling of Ref \cite{sdd} being present.
$\lambda=4$.} \label{fig10}
\end{figure}

\begin{figure}[!htb]
\includegraphics*[width=0.5\textwidth]{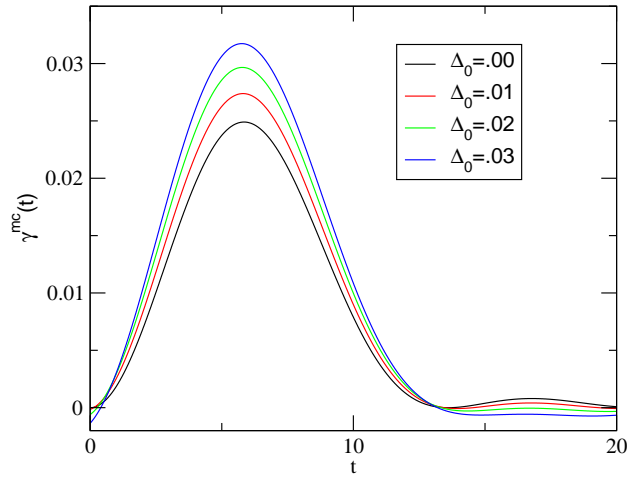}
\centering \caption{The cutoff function $\gamma^{mc}(t)$ Vs. $t$ for
the results shown in Fig \ref{fig10}.} \label{fig11}
\end{figure}

\begin{figure}[!htb]
\includegraphics*[width=0.5\textwidth]{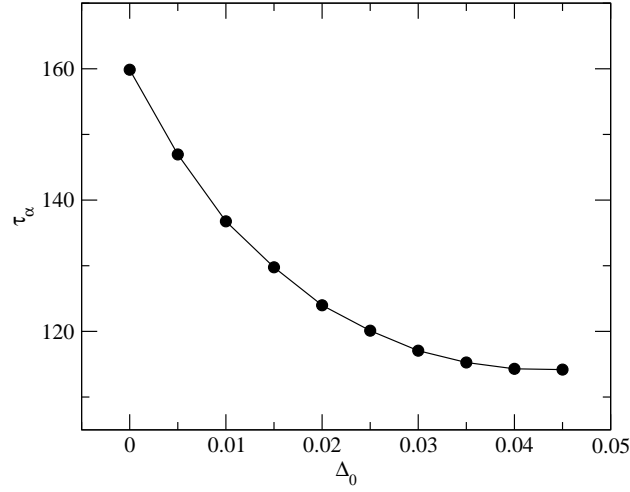}
\centering \caption{The relaxation time $\tau_\alpha$ (see text) for
the density correlation function shown in Fig \ref{fig10} vs.
diffusion constant $\Delta_0$.} \label{fig12}
\end{figure}

Finally we take the case in which both processes A and B are
included. Our primary observation here is that the scenario of
producing extremely slow dynamics due to an extra diffusive mode
$\Delta_0$ (or equivalently $\gamma_0$) for the collective density
becomes ineffective when we take in to account all the relevant
nonlinearities present in the FNH equations. Here roles of a) the
bilinear coupling of ${\bf g}$-$\delta\rho$ in the continuity
equation (\ref{cont-eqn}), and b) the pressure nonlinearity of
density fluctuations in Eqn. (\ref{mom-eom}) are taken in to account
in the theoretical analysis. It is well known that b), the pressure
non linearities give rise to the ENE transition of the MCT. The time
evolution of density correlation function in this case is shown Fig.
\ref{fig10}. The results shown here correspond to different values
of the diffusion constant $\Delta_0$ while the density dependent
parameter $\lambda$ is kept fixed at $\lambda=4$. On comparing Fig.
\ref{fig10} with Fig. \ref{fig09}, it is clear that in presence of
${\bf g}\delta\rho$ coupling in the continuity equation, the
diffusive mode $\gamma_0$ doesn't influence final time scales of
density correlation function much. The corresponding cutoff function
$\gamma^{mc}$ with the time is shown for different $\Delta_0$ in
Fig. \ref{fig11}. With increasing $\Delta_0$, for larger diffusion
constants for the extra decay mode for the density fluctuations, the
cutoff of the ENE transition becomes larger. As a result of this the
correlation function decays faster on increasing $\Delta_0$. The
relaxation times $\tau_\alpha$ corresponding to the final decay of
the density correlation function for different $\Delta_0$ values are
shown in Fig. \ref{fig12}. The nature of dependence of $\tau_\alpha$
on $\Delta_0$ also agreed with the argument that the role of
diffusion coefficient on slow dynamics process is very feeble.

\begin{figure}[!htb]
\includegraphics*[width=0.5\textwidth]{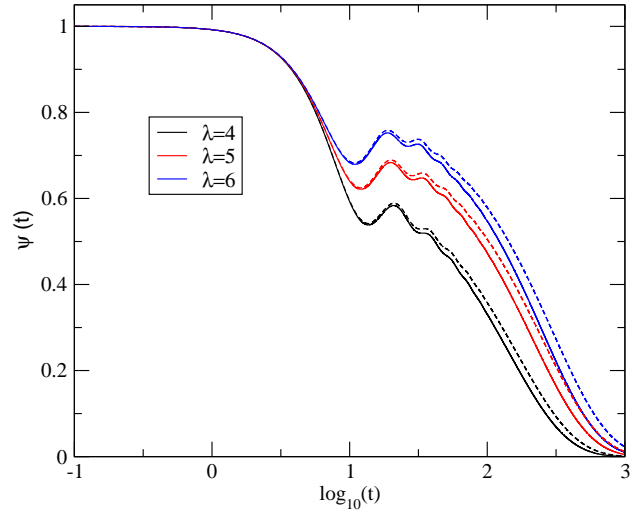}
\centering \caption{The density correlation function $\psi(t)$ vs.
log(t) with the density-current coupling of Ref. \cite{sdd} being
present and $\lambda=4$. Solid line for $\Delta_0$ treated as a
function of $\lambda$ and dashed line for $\Delta_0=0$.}
\label{fig13}
\end{figure}

\begin{figure}[!htb]
\includegraphics*[width=0.5\textwidth]{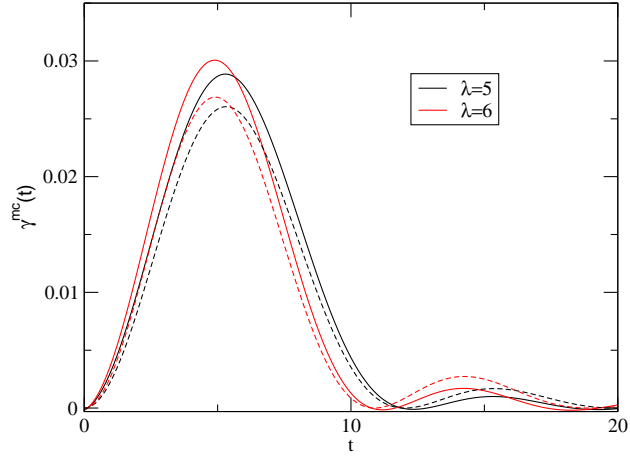}
\centering \caption{The cutoff function $\gamma^{mc}(t)$ vs. $t$
corresponding to results of Fig. \ref{fig13}. Solid and dashed lines
refer to the corresponding cases of Fig.  \ref{fig14}.}
\label{fig14}
\end{figure}

\begin{figure}[!htb]
\includegraphics*[width=0.5\textwidth]{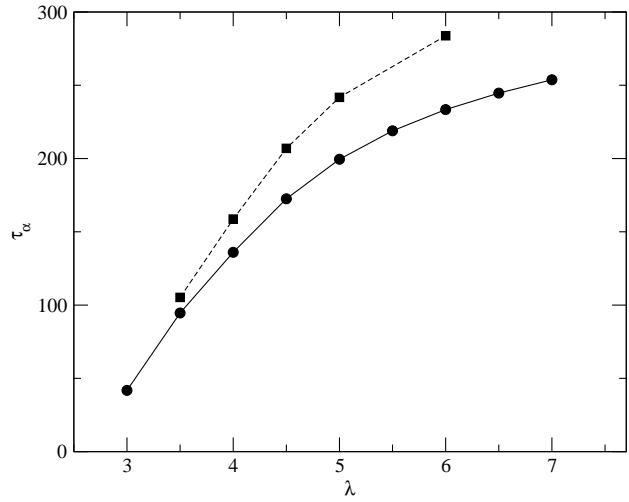}
\centering \caption{The relaxation time $\tau_\alpha$ (see text) for
the density correlation function shown in Fig \ref{fig13} vs. the
coupling constant $\lambda$.}\label{fig15}
\end{figure}

We consider here also the situation in which diffusion coefficient
$\gamma_0$ for the extra decay mode of density is dependent on the
parameter $\lambda$, {\em i.e.}, on the thermodynamic state of the
liquid. $\gamma_0$ or $\Delta_0$ is chosen the form of $\Delta_0$
\be \label{dens-del} \Delta_0(\lambda)=.01~ \exp(-(.01)\lambda) \ee
We present here the numerical results with $\Delta_0$ taken as a
function of $\lambda$. Fig. \ref{fig13} shows the behavior of
density correlation function with time for different values of
$\lambda =4$, $5$ and $6$. Solid lines. The dashed line in each case
respectively represent the corresponding correlation function
obtained by solving Eqn. (\ref{grr-t}) with $\Delta_0=0$ in Eqn.
(\ref{mem-gDM}). Similar to Fig. \ref{fig11}, we plot in Fig.
\ref{fig14} the dependence of the cutoff functions $\gamma^{mc}$ on
the parameter $\lambda$ for both of the above two cases. Solid and
dashed lines show the results for $\Delta_0$ being given by Eqn.
(\ref{dens-del}) and set to zero respectively . The correlation
function relaxes slower with increasing $\lambda$. The variation of
relaxation time $\tau_\alpha$ with the respective $\lambda$ are
shown in Fig. \ref{fig15} for the respective cases.

\section{Discussion}

The slow dynamics of a deeply supercooled liquid has been studied in
the past by various authors with models of generalized
hydrodynamics. These models, formulated in terms coupling of slow
modes in a dense liquid, are generally termed as mode coupling
theory (MCT). They can be broadly divided in the following three
groups.\\

\noindent A. The simple MCT in which an ideal ergodicity to
non-ergodicity (ENE) transition is predicted. The origin of the ENE
transition in basic MCT is the nonlinear coupling of density
fluctuations in the pressure term of the generalized Navier-Stokes
equation (\ref{mom-eom}). In a simple schematic form of this model,
beyond a critical value of the density dependent parameter $\lambda
\geq \lambda_c$, the liquid undergoes a transition from the ergodic
to non-ergodic state. The long time limit of the  density
correlation function $\psi_{\rho\rho}( t)$ is an order parameter for
this ENE transition and jumps to a nonzero value beyond the
transition \cite{gtze85,scal-jcp}. This model, in various forms have
been studied extensively in the literature for explaining data on
glassy dynamics \cite{rmp}, in particular at the initial stage of
supercooling near melting point.\\

\noindent B. Models \cite{DM,sdd,gmct} which takes in to account
wider set of nonlinear couplings of standard hydrodynamic modes like
density and currents. These nonlinearities follow from plausible
generalizations of equations of hydrodynamics to short length
scales. In these models the sharp ENE transition referred to in type
A models is smoothed off due to mechanisms which go beyond the
simple MCT. It is generally agreed that the transition of simple MCT
is cut off due to such extended MCT models. It is the initial stage
of viscous slow down in which types of model A are more relevant. It
is also known\cite{das1990} that in its present form, the fully self
consistent models of type B do not make the cutoff mechanism weak
enough to cause large increase of time scales as seen in glassy
relaxation data.\\

\noindent C. Extension of the simple MCT with presence of extra slow
modes which are introduced from phenomenological considerations
\cite{das-schilling,yeomaj,oppenheim1,biman-wolynes}. The time scale
of the structural relaxation in the supercooled state is linked to
that of the extra slow mode which assumed to be long. Thus
development of long time scales in a supercooled liquid in these
models is more built in to the formulation than spontaneously coming
out of the model. Both models B and C has the similarity that they
reduce to model A of simple MCT when the extra couplings among
hydrodynamic modes or the presence of extra slow modes are
respectively ignored.\\

With the above background, understanding the basic mechanism of
drastic slow down in structural glasses starting from the liquid
side has remained a challenge. In the present work we extend the
mathematical analysis of type B Models to combine with it the models
of types C. First, as a test, we ignore the extra nonlinearities of
type B models and consider only the pressure nonlinearity of simple
MCT together with an extra slow model like that of type C models.
The presence of slow diffusive process removes the ENE transition
and effectively determines the final decay in the supercooled liquid
as expected. The density correlation $\psi_{\rho\rho}(t)$ decays to
zero. From the Fig.\ref{fig03} it is clear that this decay occurs
increasingly slowly with decreasing $\gamma_0$. Second, we study the
case in which the extra diffusive process is ignored {\em i.e.},
$\gamma_0=0$. Now the results of Ref. \cite{sdd} model are
reproduced from our model. The present work therefore also offers a
detailed mathematical deduction of models of Ref. \cite{sdd}.
Finally, we study the combined model of B and C. Our main finding is
that in this case the decay of the density correlation is not
critically controlled by the presence of the slow diffusive mode. It
is the density current nonlinearities in the equations of
generalized hydrodynamics which produce the dominant effect on the
final decay process. The latter is determined self consistently in
terms of the correlation functions and their time derivatives
\cite{tran-d}. We ignore transverse momentum correlations
contribution to the cut off functions for simplicity and assume that
they decay much faster. The dynamics is generally slower with the
$\gamma_{DM}$ as compared to that with $\gamma_{SDD}$. In both case
however we note that, (for the schematic cases with all wave vector
dependence dropped) the ergodicity-restoring term completely wash
out the slow dynamics. This means that the latter is not small
enough to entail very slow dynamics.

The simple MCT predicts a ENE transition and in the so called
extended MCT, the role of the nonlinearities going beyond simple
coupling of density fluctuations smooths off the dynamic transition.
The key factor for producing very slow dynamics lies in how the
ergodicity restoring mechanism due to the the current-density
couplings ($1/\rho$ nonlinearity in Eqn. (\ref{cont-eqn}) for ${\bf
g}$) are getting suppressed. From a physical point of view, in the
supercooled state the effects of currents density couplings should
change in a manner so as to enhance the cage effect in the dense
liquid. Whatever we call this process, be it hopping or $1/\rho$
nonlinearities, it is indeed true that the effect of this cutoff
mechanism must be reduced to explain structural arrest from the
liquid side within MCT. The present work in fact {\em puts a
constraint of similar nature} on the phenomenological models that
have been proposed in recent literature to serve this. Going beyond
MCT, various scenarios have been proposed in this respect, including
spontaneous breakdown of ergodicity \cite{remi} and Random first
order transition \cite{wolynes-rfot} to explain this structural
arrest along the lines of Adam and Gibbs classic papers. However
coming from the liquid side, using a microscopic approach like MCT,
this still remains an open problem.

\section*{Acknowledgement}
\label{sec_ACK} NB acknowledges CSIR, India for financial support.
SPD acknowledges support under grant 2011/37P/47/BRNS.




\begin{table}[!htb]
\begin{center}
\begin{tabular}{l|cc|cc}
& $\rho$   & $g_j$   & $\hat \rho$  & $\hat {g}_j$ \\
\hline 
$\rho$ & 0 & 0 & $-\omega+ i \gamma_0 q^2$
& $q_jc_0^2$     \\
$g_i$ & 0 & 0 & $q_j$
& $-\omega\delta_{ij}+ i\tilde{ L}^0_{ij}$    \\
\hline & & & & \\
$\hat{\rho}$ & $\omega+ i \gamma_0  q^{2}$ & $- q_j $ &
$2\beta^{-1}\gamma_0 q^2\chi_{\rho\rho}$ & 0 \\ $\hat {g}_i$ & $-
q_i c_0^2$ & $\omega\delta_{ij}+i\tilde{
L}^0_{ij}$ & 0 & $2\beta^{-1}{ L}^0_{ij}$ \\
& & & & \\
\hline
\end{tabular}
\caption{The inverse of zeroth-order matrix $G_{\alpha\beta}^{0}$}
\label{table-G0}
\end{center}
\end{table}

\begin{table}[!htbp]
\begin{center}
\begin{tabular}{c|c|c}
~~~~&~~~~$\rho$~~~~&~~~~$g_j$~~~~\\
\hline ~~~~$\hat\rho$~~~~&~~~~$\omega+i q^2\gamma$~~~~
&~~~~$-q_j (1+\gamma_{\hat \rho{g}_j})$~~~~\\
~~~~&~~~~$~$~~~~&~~~~$~$~~~~\\
\hline & & \\
 ~~~~$\hat{g}_{i}$~~~~&~~~~$-q_i~c^2$~~~~
&~~~~$\omega\delta_{ij}+i L_{ij}$~~~~\\
\hline
\end{tabular}
\caption{Elements of matrix $[G^{-1}]_{\hat\alpha\beta}$ defined in
terms of the matrix ${\cal R}$.} \label{Gf-R}
\end{center}
\end{table}
\begin{table}[!htbp]
\begin{center}
\begin{tabular}{c|c|c}
~~~~&~~~~$\hat\rho$~~~~&~~~~$\hat{g}_j$~~~~\\
\hline ~$\hat\rho$~~~~&~~~~$2\beta^{-1}\gamma_0 q^2
\chi_{\rho\rho}-\Sigma_{\hat\rho\hat\rho}$ &
~~$-\Sigma_{\hat\rho\hat{g}_j}$~~~~\\
~~~~&~~~~~~~~&~~~~~~~~\\
\hline & & \\
 ~~~~$\hat{g}_i$~~~~&~~~~$-\Sigma_{\hat{g}_i\hat\rho}$
~~~~&~~~~$2\beta^{-1} L_{ij}^0-
\Sigma_{\hat{g}_i\hat{g}_j}$~~~~\\
\hline
\end{tabular}
\caption{Elements of matrix ${\cal C}_{\hat\alpha \hat\beta}$.}
\label{Gf-C}
\end{center}
\end{table}
\begin{table}[!htbp]
\begin{center}
\begin{tabular}{c|c|c}
 ~~~~&~~~~$\hat\rho$~~~~&~~~~$\hat{g}$~~~~\\
 \hline
~~~~$\rho$~~~~&~~~~$\omega+iq^2{L}$
~~~~&~~~~$q(1+\gamma_{\hat\rho {g}})$~~~~\\
~~~~&~~~~$~$~~~~&~~~~$~$~~~~\\
\hline  & & \\
 ~~~~${g}$~~~~&~~~~$q~ c^2$
~~~~&~~~~$\omega+i q^2\gamma$~~~~\\
\hline
\end{tabular}
\caption{Elements of matrix $N_{\alpha \hat\beta}$ in terms of the
renormalized transport coefficient $\gamma$.} \label{response}
\end{center}
\end{table}


\appendix

\section{Renormalization}
\label{appendix-A}

The key quantity which determined the renormalization of the
correlation function matrix from ${\bf G}_0$ to ${\bf G}$, is the
self energy matrix $\Sigma$. This is expressed in the Dyson equation
(\ref{dyson}). There are primarily two types of elements
$\Sigma_{\psi\hat{\psi}}$ and $\Sigma_{\hat{\psi}\hat{\psi}}$
respectively referred to as response and correlation type matrix
elements. Let us first consider the Dyson equation for the case in
which both indices in the matrix Eqn. (\ref{dyson}) correspond to
the un-hatted fields. In this case, we have for the two respective
terms on the right hand side,

\begin{itemize}

\item[(a)] $[{\bf G_0}^{-1}]_{\alpha\beta}=0$ which follows from the
action (\ref{MSR-action}) obtained in the MSR field theory.

\item[(b)] ${\bf \Sigma}_{\alpha\beta} =0 $ which follows from causal nature of
the response functions in MSR field theory.

\end{itemize}

\noindent We therefore obtain that the elements of the ${\bf
G}^{-1}$ matrix corresponding to the un-hatted fields, ${\left [{\bf
G}^{-1}\right ]}_{\alpha\beta}=0$. The inverse of the green function
matrix ${\bf G}$ is obtained in the following block diagonal form
\begin{equation}
\label{ga-inv} {\bf G}^{-1} = \left [
\begin{array}{cc}
\bigcirc  & {\cal R}^{\dag} \\
{\cal R} & {\cal C}\\
\end{array}
\right ]
\end{equation}
The renormalized for of the matrix ${\cal R}$ and ${\cal C}$ are
obtained in terms of the elements of the self energy matrix $\Sigma$
introduced in the Dyson equation (\ref{dyson}). Here the symbol
${\cal R}^{\dag}$ denotes the hermitian conjugate of ${\cal R}$. By
taking out the leading order wave number dependence of the
respective correlation and response self energies, we define the
various elements of $\gamma_{\psi\hat{\psi}}$ where
$\psi{\in}\{\rho,{\bf g}\}$. The renormalized elements of the matrix
${\cal R}$ and ${\cal C}$ are respectively listed in table
\ref{Gf-R} and \ref{Gf-C} .
The above matrix elements for ${\bf G}^{-1}$ are obtained from
renormalizing the corresponding zeroth order contributions in ${\bf
G}_0^{-1}$ with the appropriate self energies. The  $c$, $\gamma$
and ${L}$ respectively denote the renormalized quantities for the
sound speed $c_0$, the diffusion constant $\gamma_0$, and the
longitudinal viscosity $\tilde{L}_{ij}^0$.
\bea \label{c-ren} c^2(q,\omega)&=& c_0^2
+\gamma_{\hat {g} \rho}({\bf q},\omega)\nonumber\\
\label{D-ren} \gamma({\bf q},\omega)&=&\gamma_0+
i{\gamma_{\hat \rho \rho}({\bf q},\omega)}\nonumber\\
\label{L-ren}  L_{ij}({\bf q},\omega)&=&\tilde{L}_{ij}^0+
i\Sigma_{\hat {g}_i {g}_j}({\bf q},\omega)~~. \nonumber \eea
The contribution to $\gamma_{\hat\rho {g}}$ is taken at the lowest
order $O(q^2)$. Inverting the matrix ${\bf G}^{-1}$ having the above
structure (\ref{ga-inv}), we obtain for the correlation functions of
the physical, un-hatted field variables
\begin{equation}
 \label{eq:84a} G_{\alpha\beta}=-\sum_{\mu\nu
} G_{\alpha\hat{\mu}}C_{\hat{\mu}\hat{\nu}} G_{\hat{\nu}\beta}
\end{equation}
where Greek letter subscripts take values $\rho ,{\bf g}$, and the
self energy matrix $\Sigma_{\hat{\mu}\hat{\nu}}$ determine the
corresponding elements of the matrix ${\cal C}$.
From the set of equations denoted by (\ref{dyson}) we also obtain
that the response functions $G_{\alpha\hat{\mu}}$. Here we make use
of the functional identity,
\begin{eqnarray}
\int  D(\psi)\frac{\delta}{\delta\psi_{\hat \alpha}(1)}\left[
\psi_{\beta}(2) {e}^{-{\cal A}[\psi]}\right]=0 \nonumber\
\end{eqnarray}
where $D(\psi)$ denotes the functional integral with the fields
$\{\psi\}$ and ${\cal A}$ is the MSR action ( for example see Eqn.
\ref{MSR-action}) ) with respect to which averaged are obtained.
Using this identity we obtain
\begin{equation} \label{dyson-resp}
\left[(G_{0}^{-1})_{\hat{\alpha}\mu}(13)
-\Sigma_{\hat{\alpha}\mu}(13) \right]G_{\mu\hat{\beta}}(32)= \delta
(12)\delta_{\hat{\alpha}\hat{\beta}}~~.
\end{equation}
The self energies $\Sigma_{\hat{\alpha}\mu}$ are expressed in a
perturbation theory in terms of the two-point correlation and
response functions. Now inverting the matrix ${\cal R}$ given in
table \ref{Gf-R}, we obtain the response function in the following
form.
\begin{equation}
\label{resp-funa} G_{\alpha
\hat{\mu}}=\frac{N_{\alpha\hat{\mu}}}{\cal D}
\end{equation}
where the matrix $N$ is given in table \ref{response}. The
determinant ${\cal D}$ in the denominator of Eqn. (\ref{resp-funa})
is given by.
\begin{equation}
\label{detrmina}{\cal D}= (\omega+ iq^{2}\gamma)~
(\omega+iq^2{L})-q^2c^2~~.
\end{equation}
$L(q,\omega)$ is the renormalized longitudinal viscosity obtained in
terms of the self energy matrix:
\be \label{ren-longa}  L({\bf q},\omega)=\tilde{L}_0+ i\gamma_{\hat
{g} {g}}({\bf q},\omega)~~. \nonumber \ee
The bare longitudinal viscosity is obtained as
$L_0=(\zeta_0+4\eta_0/3)$ and the corresponding kinematic viscosity
$\tilde{L}_0$ in Eqn. (\ref{ren-longa}) is denoted as
$\tilde{L}_0=L_0/\rho_0$.

The symmetries of the vector field ${\bf g}$ and the scaler field
$\rho$ require that the vertices having the corresponding MSR hatted
field $\hat{g}_i$ or $\hat{\rho}$ are associated with a factor of
$q$, due to the total derivatives present in the nonlinearities in
the respective FNH equations. Using this, we first estimate the
leading order wave vector dependence of the various self energy
functions. The correlation self energy, elements between two hatted
fields are obtained as
\bea \label{cor-se1} \Sigma_{\hat {g}_i \hat {g}_j}({\bf
q},\omega)&=&
- q^2\gamma_{\hat {g}_i \hat {g}_j}(0,\omega)\\
\label{cor-se2} \Sigma_{\hat \rho \hat \rho}({\bf q},\omega)&=&-
q^2\gamma_{\hat \rho \hat \rho} ( 0,\omega)\\
\label{cor-se3} \Sigma_{\hat \rho \hat {g}_i}({\bf q},\omega)&=&-
q^3 \gamma_{\hat \rho \hat {g}_i} ( 0,\omega) \eea
On the other hand the response self energy elements between hatted
and un-hatted fields $\Sigma_{\psi\hat{\psi}}$ generally satisfies
the condition
\be \label{resp-self} \Sigma_{\psi\hat{\psi}} ({\bf
q},\omega)=-\Sigma^*_{\hat{\psi}{\psi}}({\bf q},\omega) \ee
The leading order behavior of the response self energy elements are
obtained as
\bea \label{res-se1}\Sigma_{\hat {g}_i {g}_j}({\bf q},\omega)&=&-
i  q^2\gamma_{\hat {g}_i {g}_j}( 0,\omega)\\
\label{res-se2} \Sigma_{\hat \rho \rho}({\bf q},\omega)&=&- i
 q^2 \gamma_{\hat\rho \rho}( 0,\omega)
\\
\label{res-se3} \Sigma_{\hat{g}_i \rho}({\bf q},\omega)&=&-i q^3
\gamma_{\hat {g}_i \rho}( 0,\omega)
\\
\label{res-se4} \Sigma_{\hat \rho {g}_i}({\bf q},\omega)&=&-i q^3
\gamma_{\hat \rho {g}_i}( 0,\omega) \eea
We find that at the one loop order, for the self energies on the
right hand side of Eqns. (\ref{res-se3})-(\ref{res-se4}), the $O(q)$
contribution is vanishing. In a similar way the total contribution
to the self energy $\gamma_{\hat{\rho}{g}}$ from the one loop
diagrams involving the nonlinearity in the continuity equation is of
$O(q^2)$. The latter therefore is ignored compared to 1 in the
matrix element $N_{\rho\hat{g}}$ shown in table \ref{response}.
These diagrams are explicitly shown at the end of this section.
Using the above forms for the self energies in the fluctuation
dissipation relations (\ref{FDR-g})-(\ref{FDR-r}) and doing a
leading order analysis in the hydrodynamic limit, we obtain a set of
useful relations between the correlation and response self energies.
These relations are essential in demonstrating that the theory is
renormalizable in terms of a redefined quantities presented in Eqns.
(\ref{c-ren})-(\ref{L-ren}), in terms of the self energy functions.
Doing a standard analysis \cite{DM} of the expression (\ref{eq:84a})
for the correlation functions, the following self-energy relations
are obtained in the hydrodynamic limit
\bea \gamma_{\hat{g}\hat{g}}&=&2~\beta^{-1}\rho_{0}
\gamma'_{\hat{g} g}\\
\gamma_{\hat\rho\hat\rho}&=&-2~\beta^{-1}\frac{\rho_{0}}{ c^2_0}
\gamma^{'}_{\hat \rho\rho }~~., \eea
where the prime indicates the real part of the corresponding self
energy element. The renormalized viscosity is given by either in
terms of the response or the correlation self energies. We write \be
L=L_0+\frac{\beta}{2}\gamma_{\hat{g}\hat{g}}~~.\ee
Similarly the renormalized diffusion coefficient is given by \be
\gamma=\gamma_0+\frac{\beta}{2}\frac{ c^2_0}{\rho_{0}}
\gamma_{\hat\rho\hat\rho}\ee

\subsection{One loop contribution for $\Sigma_{\hat{\rho}g}$}

\begin{figure}[!htb]
\includegraphics*[width=0.5\textwidth]{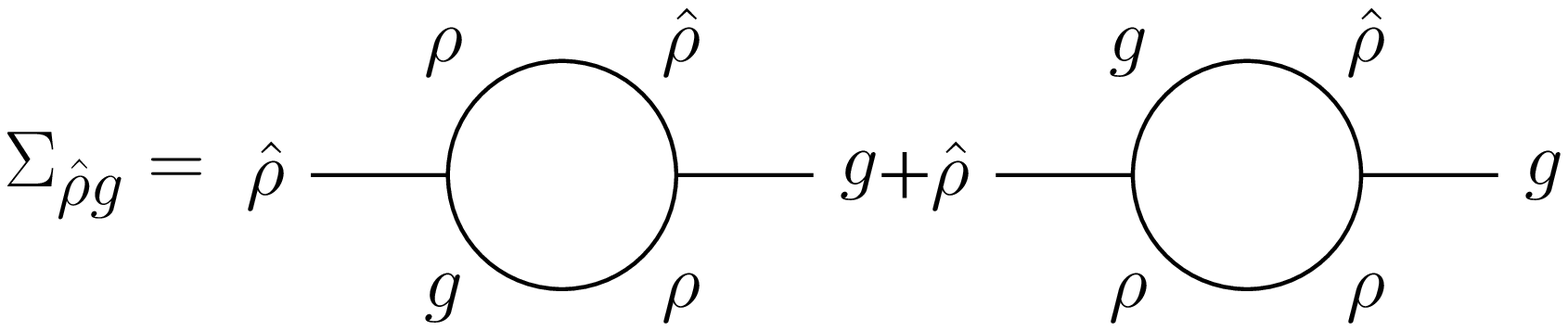}
\centering \caption{One loop contributions to
$\Sigma_{\hat{\rho}g}$.}\label{fig16}
\end{figure}

The contribution to the first diagram ( see Fig. \ref{fig16} ) for
$\Sigma_{\hat{\rho}{g}}$ as shown in Fig. \ref{fig16} is obtained as
\be \Sigma_{\hat{\rho}
g}^{(1)}(q,t)=\frac{q}{\rho_0^2}\int\frac{d{\bf k}}{(2\pi)^3}
u(q-uk) ~G_{\rho\hat{\rho}}(q-k,t)~G_{ g \rho}(k,t)~~. \ee
$u$ is the cosine of the angle between ${\bf q}$ and ${\bf k}$. For
the second diagram shown in the same figure we obtain
\be \Sigma_{\hat{\rho}{
g}}^{(2)}(q,t)=\frac{q}{\rho_0^2}\int\frac{d{\bf k}}{(2\pi)^3} u^2 k
~G_{\rho\rho}(q-k,t)~G_{g \hat{\rho}}(k,t). \ee
We use the Fluctuation-dissipation relation
\be i\Theta(t)G_{\beta\rho}(q,t)=
\beta^{-1}\chi_{\rho\rho}G_{\beta\hat{\rho}}(q,t), \ee
where $\Theta(t)$ is the Heaveside step function. We obtain for the
total contribution from the sum of the two diagrams as,
\bea \Sigma_{\hat{\rho}{ g}}(q,t)&=&\Sigma_{\hat{\rho}g}^{(1)}(q,t)
+\Sigma_{\hat{\rho} g}^{(2)}(q,t)\nonumber\\
&=&i\Theta(t)~q^2\frac{\beta c^2}{\rho_0^2}\int\frac{d{\bf
k}}{(2\pi)^3}\Big( u~G_{\rho\rho}(q-k,t)~G_{ g \rho}(k,t)\Big)~~.
\eea


\end{document}